# Development of SQUID Array Amplifiers for the LiteBIRD CMB Satellite

S. T. P. Boyd (ORCID 0000-0003-1526-3690) and Tijmen de Haan (ORCID 0000-0001-5105-9473)

*Abstract*—LiteBIRD is an upcoming JAXA-led mission that aims to measure primordial gravitational waves in the B-mode polarization of the cosmic microwave background. It is set to launch in 2032. The LiteBIRD detector array consists of around 5000 TES detectors which are read out using digital frequency multiplexing over a bandwidth of 1-6 MHz. The multiplexing factor ranges from 58x to 68x. We are presently developing single-stage SQUID array amplifiers for LiteBIRD readout. Due to the reduced complexity and cost, and greater heritage from ground-based experiments such as the South Pole Telescope and Simons Array, single-stage SQUID array amplification is preferable for the first-stage amplification, as long as it can meet the requirements. The LiteBIRD single-stage SQUID Array is required to have high transimpedance amplification while maintaining a low input inductance and low dynamic resistance. In addition, the input-referred current noise must be very low, and the power dissipation must remain below about 100 nW. These requirements have non-trivial interactions. To maximize performance within these requirements we have performed lumped-element SQUID simulation. We find that by optimizing SQUID internal damping elements and inductive loading, good single-stage SQUID array performance can be obtained for LiteBIRD, including significant engineering margin.

## I. INTRODUCTION

IN this report we describe a design study for the development of a single-stage Superconducting Quantum Interference Device (SQUID) Array Amplifier (SAA) [1][2] for the readout system of the upcoming JAXA-led LiteBIRD satellite mission. LiteBIRD, planned for launch in 2032, will probe cosmic inflation by surveying polarization of the Cosmic Microwave Background (CMB) [3].

The LiteBIRD detector array will consist of approximately 5000 Transition-Edge Sensor (TES) bolometers [4][5], which will be read out using a digital frequency multiplexing (dfMux) scheme [6] over a bandwidth of 1–6 MHz. The

This work was supported through a NASA Established Program to Stimulate Competitive Research (EPSCoR) Rapid Response Research (R3) grant award number NM-80NSSC23M0152.

(Corresponding author: S. T. P. Boyd).
S. T. P. Boyd is with the University of New Mexico, Albuquerque, NM 87131 USA (e-mail: stpboyd@unm.edu).
Tijmen de Haan is with the Institute of Particle and Nuclear Studies (IPNS), and with the International Center for Quantum-field Measurement Systems for Studies of the Universe and Particles (QUP-WPI), both within the High Energy Accelerator Research Organization (KEK), Tsukuba, Ibaraki 305-0801, Japan (e-mail: tijmen@post.kek.jp).

Color versions of one or more of the figures in this article are available online at http://ieeexplore.ieee.org

multiplexing factor will range from 58 to 68 LC resonators per SAA, enabling efficient readout of the large detector array. Achieving excellent performance while minimizing risk requires striking a balance between leveraging heritage technologies from prior experiments such as SPT-3G [6][7][8], and Simons Array [9], and incorporating new innovations based on the lessons learned from these ground-based deployments.

For LiteBIRD, one critical modification is the relocation of the SAA from the conventional 4 Kelvin stage to a sub-Kelvin stage closer to the focal plane. This significantly reduces parasitic inductance of the SAA input circuit—a known limitation in previous systems that adversely impacted stability, linearity, and crosstalk.

The nominal base temperature for the SAA is 400 mK, providing significant noise reduction compared to 4.2 K, but also necessitating a low power-dissipation design that simultaneously meets stringent electrical performance criteria.

Two-stage SAAs, such as Kiviranta's designs for the X-ray Integral Field Unit of the Athena space observatory [10] and SPICA SAFARI [11], are also under consideration for LiteBIRD. However, due to the reduced complexity and cost, and the greater heritage from the South Pole Telescope and Simons Array, a single-stage SAA is preferable if it can meet the requirements.

## II. REQUIREMENTS

The single-stage SAA is responsible for amplifying the weak current signals generated by the TES bolometers and transmitting them to room temperature for detection by a low-noise amplifier (LNA). The following factors place requirements on the SAA:

1. The combination of the dynamic output resistance, $R_\text{dyn}$, of the SAA and the stray capacitance in the wiring to room temperature forms an unwanted RC filter, attenuating the output voltage before reaching the LNA. This attenuation can be mitigated most simply by minimizing $R_\text{dyn}$. Alternatively, a large output voltage swing can be generated via high transimpedance $Z$, but this creates a requirement for equalization at a later stage.

2. The input-referred current noise $i_n$ must be sufficiently low to ensure that the readout system does not increase the intrinsic white noise level of the TES by more than 10%.

3. The nonzero input inductance $L_{input}$ of the SAA creates inefficiency in the Digital Active Nulling (DAN) [12] loop, increasing the noise contribution of both the SAA and LNA [13]. This can be overcome by increasing TES resistance, decreasing LNA and SAA noise, high $Z$, or low $L_{input}$.
4. The power dissipation at the SAA stage is critical, as the LiteBIRD satellite relies on a single low-power adiabatic demagnetization refrigerator with limited cooling power on the order of tens of µW for all SQUIDs. The SAA power dissipation $P$ should be as far below the 100 nW nominal maximum as possible.

Given the complex interdependencies among the SAA parameters $Z$, $L_{input}$, $R_{dyn}$ and $i_n$, optimizing the SAA requires considering their collective effect, rather than isolating individual requirements. For this reason, the performance of an SAA for LiteBIRD is actually best determined via a LiteBIRD-developed software tool [14].

To accurately model the noise performance, we employ the circuit model developed by Montgomery *et al.* [8] to represent the impedances within the readout system. We assume a 0.7 Ω operating resistance for the TES bolometers, and incorporate the relevant noise sources including the synthesizer chain noise, the input-referred SAA noise, and the demodulation chain noise (including the effects of the first-stage low-noise amplifier). We utilize NGSPICE [15] via PySPICE [16] to calculate the necessary transfer functions, which allows us to determine NEI$_{global}$, the total noise referred back to the TES. LiteBIRD imposes a noise requirement of < 8.0 pA/√Hz for a TES with an LC resonator centered at 4 MHz. The SAA influences this overall noise figure NEI$_{global}$ in several key ways. The input-referred SAA noise $i_n$ contributes directly to the total noise. The SAA's input inductance $L_{input}$ impacts the nulling efficiency, often described as "current sharing" [9][17]. The dynamic output resistance of the SAA, $R_{dyn}$, affects signal loss in the wiring harness leading to room temperature. Finally, the SAA's transimpedance $Z$ influences the TES-referred noise by scaling down the contribution from the demodulation chain noise.

The goals of this project were, first, to develop SAA designs achieving NEI$_{global} \leq 8.0$ pA/√Hz and dissipated power $P \leq 100$ nW, and second, to drive these parameters as far below their upper bounds as possible, to maximize design margin for the satellite experiment.

### III. TECHNICAL APPROACH

To maximize engineering flexibility within these tight requirements, we performed lumped-element modeling of the SQUIDs that make up the SAA[18][19]. This approach allows us to model the complex nonlinear interactions of the Josephson junctions with the SQUID input transformers and internal damping elements of interest [20][21][22][23][24] (see Fig. 1). It also allows assessment of SQUID properties versus both current and flux biases.

We have previously hand-coded lumped-element SQUID simulations in MATLAB. This provided complete control over the numerical integration and thermal noise inputs. However, the necessity to manually set up the differential equations for each new circuit topology, and the lack of a graphical user interface, was too cumbersome for this project.

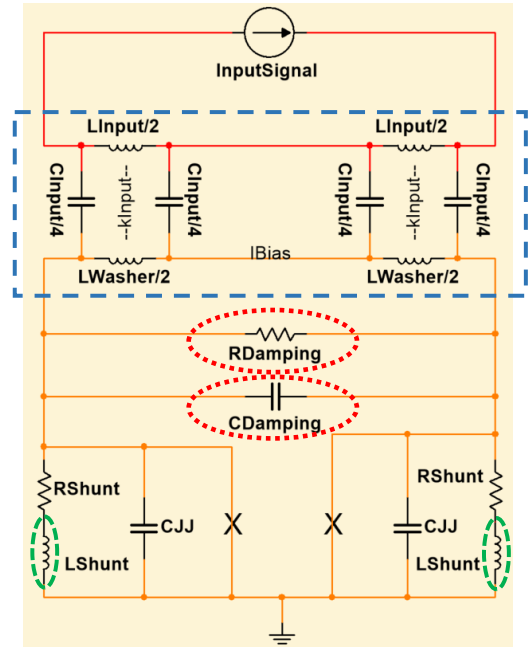

**Fig. 1.** Simplified lumped-element SQUID schematic. "Damping" elements are circled with dotted lines. Shunt inductive loading is circled with dashed lines. Input transformer is in dashed box. See text for discussion.

To address these deficiencies, we tested several SPICE-based circuit simulation packages: Multisim [25], NGSPICE [26], and LTspice [27]. Josephson junctions can be modeled using controlled sources in SPICE, as discussed in [28]. Initial testing found Multisim and NGSPICE had insufficiently robust simulation convergence, once we started increasing the number of circuit elements within the SQUID.

LTspice provided more robust convergence when modeling SQUIDs with multiple-element input coils and internal damping elements, very rarely terminating a simulation early. LTspice also supports file-driven piecewise-linear arbitrary voltage and current sources (as does Multisim). This allows easy modeling of resistor thermal noise by generating bandwidth-limited white noise input files. LTspice also has several significant practical advantages: 1) it is free to use, 2) it has a large online user community that makes it easy to work with, 3) it is supported by a major electronics company, and 4) it has a straightforward command-line interface that allows easy control via software.

However, we did find one major caveat for getting good SQUID simulations with LTspice: the user-specified maximum timestep is only respected when working with the slower ASCII-format .raw file output. The default simulation parameters, which use the binary .raw file, try to power through simulations quickly using prediction algorithms for step size, and these don't work well for the Josephson oscillations. Fully-consistent timestep-controlled resolution of the Josephson oscillations was only obtained by using the ASCII .raw file output option.

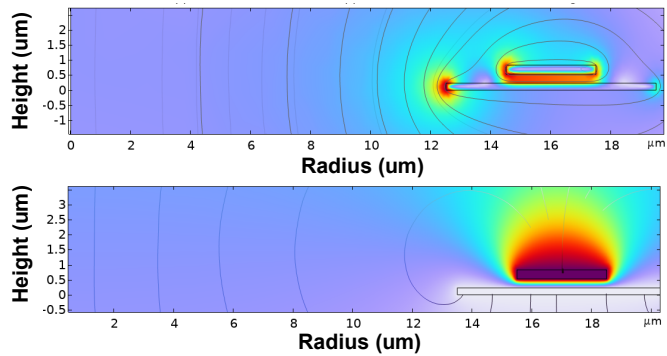

**Fig. 2.** Finite-element modeling of inductances, mutual inductance, and capacitance, for the 1-turn 24 µA/$\phi_0$ input transformer. (Upper) magnetic field. (Lower) electric potential.

We ran two types of simulations. Firstly, for parameter exploration and "tuning" of SQUIDs, we ran noiseless (0 K) "$V$-$\phi$" simulations. In particular, 0 K parametric sweeps were performed to assess internal damping parameters in the presence of input transformers of interest. These simulations generated $V$-$\phi$ curves using a current ramp through the transformer input. This simulated a triangle wave input with an 800 ns period, so with a fundamental frequency in the few-MHz LiteBIRD dfMux frequency band.

Secondly, once parameters were somewhat narrowed by the $V$-$\phi$ simulations, a more time-consuming series of 0.4 K "flux" simulations, each at a fixed flux and bias current operating point, were performed. The simulation period was long enough for transients to die away and for noise power spectra of the steady state to reach frequencies down to a few MHz. White voltage noise was determined by fitting to the portion of the noise power spectrum below a cutoff frequency, typically 1 GHz for this study.

In the "flux" simulations, MATLAB controlling code was used to set up and run the parametric sweeps, including generation of band-limited white noise appropriate for the resistor values and temperature, modification of the LTspice input files for all swept parameters and noise input files, simulation execution, and reduction of the data from the output .raw files.

Each "flux" simulation was performed 3 times: once with the baseline parameters, once with the bias current slightly increased so $R_{dyn}$ could be determined, and once with the bias flux slightly increased so $dV/d\phi$ and $Z$ could be determined. With $dV/d\phi$ in hand, $i_n$ could then be determined from the fitted white voltage noise. SQUID mean power $P$ was determined as the product of the bias current and the mean voltage across the SQUID. $L_{input}$ was from the input transformer calculations described below.

Lastly, to assess LiteBIRD performance, the Python software tool provided by LiteBIRD [14] was also run by the MATLAB controlling code, at the end of the "flux" simulation. Tables of $Z$, $i_n$, $L_{input}$, and $R_{dyn}$ for the SAA, versus the number of SQUIDs in the SAA, $N_{SQUIDs}$, were provided as input for the Python code. This yielded NEI$_{global}$ versus $N_{SQUIDs}$ for the SQUID configuration and operating point under study. To roughly account for the input inductance parasitics in the SAA, $L_{input}$ for the array was assumed to be $N_{SQUIDs} \times 1.1$ of the SQUID $L_{input}$.

TABLE I
INPUT TRANSFORMER PARAMETERS

| turns | $1/M_i$ (µA/$\phi_0$) | $L_{washer}$ (pH) | $L_{input}$ (pH) | $C_{input}$ (pF) | $k_{input}$ |
|---|---|---|---|---|---|
| 1 (fig. 8) | 8 | 269 | 300 | 0.269 | 0.909 |
| 1 | 8 | 269 | 300 | 0.269 | 0.909 |
| 3 | 8 | 89.1 | 856 | 0.504 | 0.936 |
| 1 | 24 | 91.2 | 106 | 0.126 | 0.879 |
| 3 | 24 | 30.1 | 312 | 0.308 | 0.890 |

For the NEI$_{global}$ plots shown in this report, a maximum allowed dissipated power in the SAA, $P_{max}$, is specified for each curve or set of curves. Each point in a curve shows the minimum value of NEI$_{global}$ that could be obtained for that SQUID configuration and operating point, within the range of $N_{SQUIDs}$ considered, and with the stipulation that SAA dissipated power for that SQUID configuration and operating point be less than $P_{max}$.

## IV. INPUT TRANSFORMERS

For this exploratory project, LiteBIRD performance was calculated for SQUIDs with 5 styles of input transformer that could be suitable for an SAA: 3-turn and 1-turn input coils, each with both a typical 24 µA/$\phi_0$ and the LiteBIRD maximum allowed 8 µA/$\phi_0$ input coupling, plus a 1-turn figure-8 input transformer at 8 µA/$\phi_0$. Round input coils were assumed, allowing quick calculation of inductances and capacitances in 2D axisymmetric COMSOL models [29] (Fig. 2).

Typical fabrication parameters of our industrial collaborator STAR Cryoelectronics (STARCryo) [30] were used: 230 nm/300 nm/300 nm thickness for Nb/SiO$_2$/Nb, SiO$_2$ dielectric constant = 5, Josephson junction critical current $I_c$ = 12.6 µA and junction capacitance $C_{JJ}$ = 600 fF. Layout geometry assumed standard STARCryo contact lithography tolerances: 3 µm-wide input coil traces with 2 µm spacing, atop washers with 2 µm radial extension past the inner and outer radii of the input coil turns.

Calculated parameters of the input transformers are summarized in Table I. The 1-turn 8 µA/$\phi_0$ figure-8 coil differs from the non-figure-8 coil only in the number of transmission-line segments per turn and the wiring that connects them, in the lumped element model.

We also studied selected cases using 700 nm SiO$_2$ thickness to lower $C_{input}$, but saw little difference in the single-SQUID studies of this project. However, given the propensity of SAAs toward instability and resonances, we plan to also include the 700 nm SiO$_2$ thickness in the initial SAA fabrications.

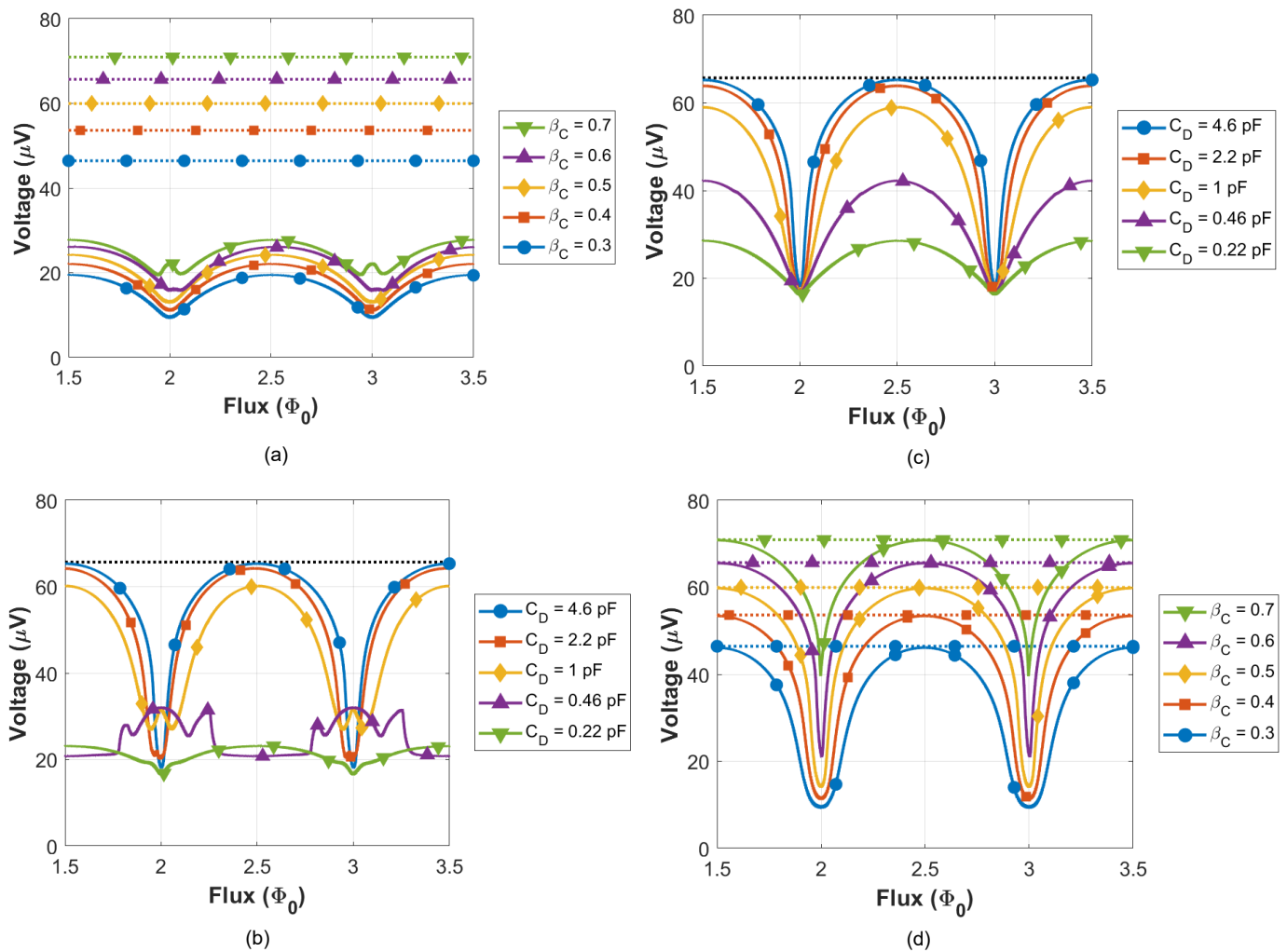

**Fig. 3.** $V$-$\phi$ curves summarizing simple heuristic "tuning" procedure for the 8 μA/$\phi_0$ figure-8 1-turn input transformer. The dotted horizontal lines show $I_c R_{shunt}$, which is the maximum possible peak height for each $V$-$\phi$ curve. (a) baseline SQUID, varying $\beta_c$. (b) sweeping $C_{damping}$ at $\beta_c = 0.6$ to raise the peaks to $I_c R_{shunt}$. Note that "$C_{damping}$" has been abbreviated to "$C_D$" here and in (c). (c) adjusting $R_{damping}$ to stabilize the $V$-$\phi$ curves. (d) re-check $\beta_c$ ladder for "tuned" SQUID, ($C_{damping} = 10$ pF, $R_{damping} = 37$ Ω). This procedure maximizes $V$-$\phi$ amplitude, stabilizes $V$-$\phi$ curves, and increases maximum $dV/d\phi$ at $\beta_c = 0.6$ from ~50 μV/$\phi_0$ to ~700 μV/$\phi_0$. Tuning procedures of this type were performed for each input transformer considered.

## V. Design Strategy

To the extent that it would be compatible with excellent LiteBIRD performance, the optimal SAA design would follow the SAA design approach introduced by Drung and co-workers at PTB [31]. This design combines narrow traces for flux expulsion with single-turn figure-8 input transformers formed by input coil traces distributed across closely-spaced narrow-line figure-8 SQUID washers. This approach has several key benefits: 1) excellent resistance to flux trapping when cooling down in ambient magnetic field, 2) high immunity to thermal noise currents in nearby normal metal, and 3) the distribution of the input coil traces breaks up the transmission-line segments of the input transformers into multiple short lengths, mixing up their order and raising their resonance frequencies, which in principle decreases possible sources of instability in the SAA.

However, considering the PTB SAA approach for LiteBIRD creates an interesting question in the context of the LiteBIRD requirements. $L_{input}$ rises as the square of the number of turns in the input coil. The single-turn input transformer of the PTB approach will minimize $L_{input}$, but at the cost of increased $L_{washer}$ (see Table I). Larger $L_{washer}$ normally degrades SQUID performance, but this can be counteracted by resistive [23][24] and/or capacitive [20][21][22] "damping" elements (see Fig. 1 schematic). Thus a key goal of this project was to determine if the inclusion of such damping elements can facilitate good SQUID performance for LiteBIRD.

## VI. Heuristic SQUID Tuning Procedure

We used a simple heuristic "tuning" procedure based on 0 K $V$-$\phi$ simulations to select values for $C_{damping}$ and $R_{damping}$. Fig. 3 summarizes this procedure, which worked well for this project.

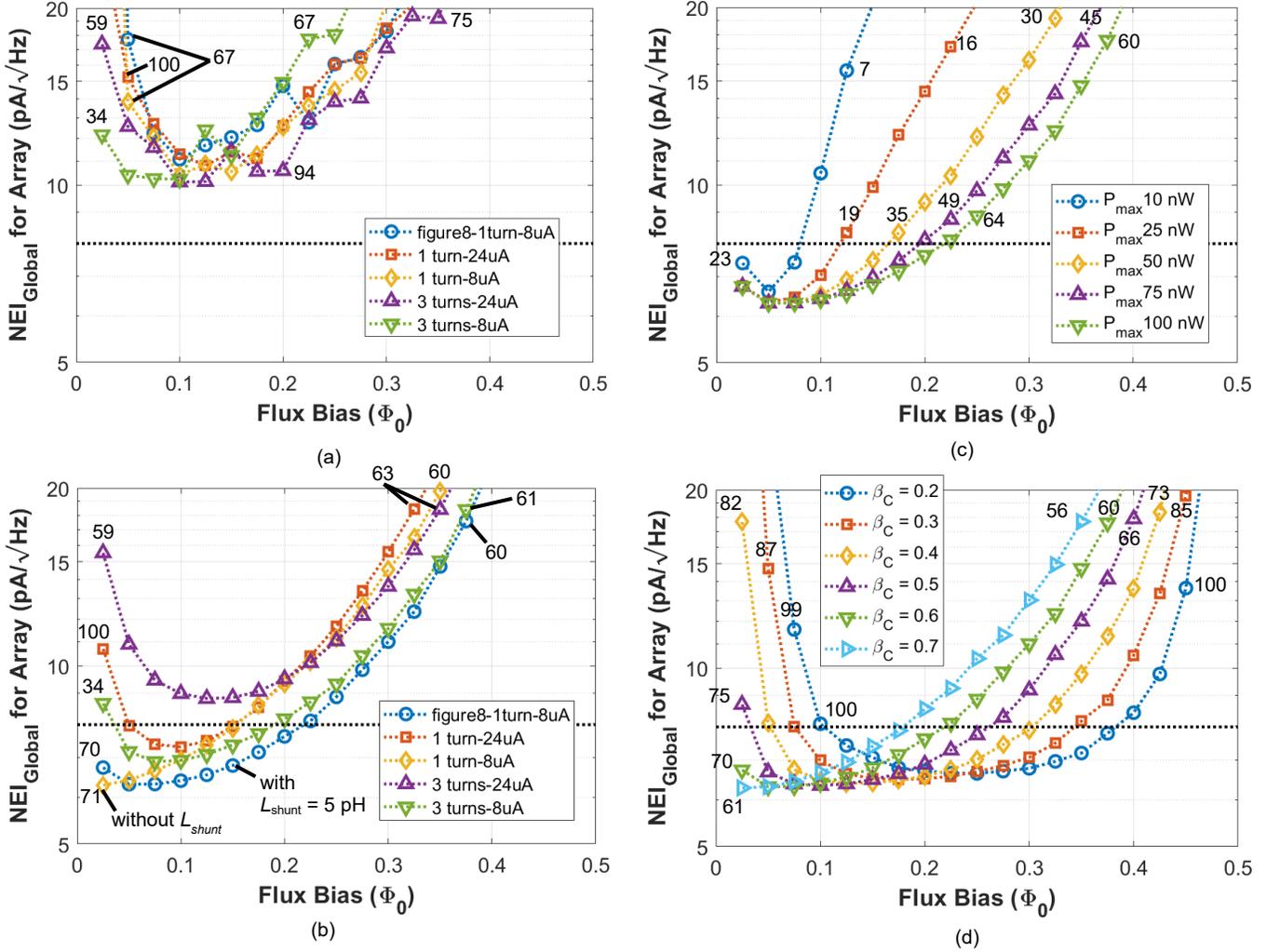

**Fig. 4.** LiteBIRD performance metric NEI$_{global}$ versus flux operating point for 4 different studies. Flux bias $0\phi_0$ ($0.5\phi_0$) is the valley (peak) of the SQUID $V$-$\phi$ curve. The horizontal dotted line in each plot is the maximum allowed NEI$_{global}$. Each plotted point is the lowest NEI$_{global}$ achievable among SAA configurations with $N_{SQUIDs}$ in the range 1 – 100, and with the SAA total dissipated power $P \leq P_{max}$, for specified $P_{max}$. The numbers near selected data points show the optimal $N_{SQUIDs}$ values for those points. Left-side plots (a) & (b): Performance of the 5 different SQUID input transformers and the impact of the heuristic "tuning" procedure, for parameters $P_{max} = 100$ nW, $\beta_c = 0.6$, and bias current $1.02 \times 2I_c$. (a) none of the "baseline" SQUIDs without damping elements achieved acceptable performance for LiteBIRD. (b) All of the "tuned" SQUIDs with optimized damping elements reach below the NEI$_{global}$ cutoff over some range of flux bias, except for the 3-turn / 24 μA/$\phi_0$ input transformer. Right-side plots (c) & (d): LiteBIRD performance metric NEI$_{global}$ versus flux operating point, for the 8 μA/$\phi_0$ figure-8 single-turn input transformer only, "tuned" with $C_{damping} = 10$ pF, $R_{damping} = 37$ Ω (see Fig. 3), and shunt inductive loading $L_{shunt} = 5$ pH (see text). (c) Performance versus $P_{max}$ at $\beta_c = 0.6$, showing decreasing acceptable flux bias range with decreasing power. (d) Performance versus $\beta_c$ for $P_{max} = 100$ nW, showing increasing acceptable flux bias range with decreasing $\beta_c$.

We first checked the $V$-$\phi$ curves for the "baseline" SQUID, with no damping elements, for a given input transformer. An example ladder of $V$-$\phi$ curves versus Stewart-McCumber parameter $\beta_c$ for the 1-turn figure-8 8 μA/$\phi_0$ input transformer is shown in Fig. 3 (a). The horizontal dashed lines indicate $I_cR_{shunt}$, the maximum possible $V$-$\phi$ peak height for each curve. We see that the $V$-$\phi$ peak heights are well below their $I_cR_{shunt}$ maximum values, because the input transformer impedes charge exchange between the junctions. We also see that 0.5 is the largest $\beta_c$ with well-formed $V$-$\phi$ curves, reflecting the resonance structure.

The second step is to assess the efficacy of $C_{damping}$ in creating a low-impedance charge-exchange path to raise the $V$-$\phi$ peaks towards $I_cR_{shunt}$, as shown in Fig. 3 (b). We do this at a relatively high $\beta_c$, where the lower damping will make $V$-$\phi$ curve defects more pronounced. We see that $C_{damping}$ greater than about 3 pF has peaks close to $I_cR_{shunt}$, but there are badly-formed $V$-$\phi$ curves at 1 pF and below.

In the third step we choose a $C_{damping}$ value and explore $R_{damping}$ values to stabilize the $V$-$\phi$ curves. We prefer larger $R_{damping}$ to minimize the thermal noise contribution. Fig. 3 (c) shows the $V$-$\phi$ curves for $R_{damping}$ = 37 Ω, which contributes only about 0.1 $\mu\phi_0/\sqrt{Hz}$ at 0.4 K.

In the fourth step, Fig. 3 (d), we check the $\beta_c$ ladder for our chosen $C_{damping}$ = 10 pF and $R_{damping}$ = 37 Ω. Compare Fig. 3 (d) to (a). At the cost of a small increase in noise from $R_{damping}$ (which could be omitted for this SQUID if necessary, if $C_{damping}$ > about 5 pF), we have increased the maximum transfer function $dV/d\phi$ from about 50 $\mu V/\phi_0$ to about 700 $\mu V/\phi_0$.

Simple tuning procedures of this type, with minor variations, were performed for each input transformer considered. The results discussed below used the optimal damping component values for each different input transformer.

## VII. MODELING RESULTS AND DISCUSSION

Initially, simulations with both single-string and parallel-bank SAAs were performed, to check the relative importance of $R_{dyn}$. We found that, within the NEI$_{global}$ assessment, single-string SQUID arrays always provided the best estimated LiteBIRD performance at a given power dissipation. All results presented here are for single-string SAAs.

As shown in Fig. 4 (a), none of the modeled SQUID configurations without tuned damping elements were able to meet the minimum performance needed for LiteBIRD.

Of the SQUID configurations with tuned damping elements, all but one reach below the NEI$_{global}$ cutoff line over some range of flux bias, as shown in Fig. 4 (b). The weakest performers are the two with weaker input coupling 24 $\mu A/\phi_0$. The lowest values of NEI$_{global}$, near 6.3 pA/$\sqrt{Hz}$, are attained by the two single-turn 8 $\mu A/\phi_0$ configurations.

The difference in shape between the two single-turn 8 $\mu A/\phi_0$ curves in Fig. 4 (b) results from the addition of $L_{shunt}$ = 5 pH of inductive loading to the SQUID shunt resistors of the figure-8 coil (see Fig. 1). Early investigation of the utility of $L_{shunt}$ by one of us (SB) has shown benefit for pulling the valley bottoms of $V$-$\phi$ curves downwards, and for suppressing hysteresis in the RCSJ model [32][33]. A separate publication on this topic is planned after a thorough investigation is completed. For the present LiteBIRD study, this inductive loading adds engineering margin by significantly increasing the working area under the NEI$_{global}$ cutoff.

Fig. 4 (c) and (d) show the NEI$_{global}$ behavior for the "tuned" figure-8, 8 $\mu A/\phi_0$ configuration, including the inductive loading of the shunt, versus $P_{max}$ (c) and $\beta_c$ (d). Operation at significantly reduced SAA power is possible, at the cost of reduced working area under the NEI$_{global}$ cutoff line. Operation at lower $\beta_c$ broadens the usable flux bias range with only a slight rise in NEI$_{global}$, and centers it about 0.25$\phi_0$, which may simplify the SQUID tuning process.

## ACKNOWLEDGMENT

SB is grateful to R. Cantor, M. Kiviranta, and J. Beyer for discussions on SQUIDs.